\documentclass{aa}  

\usepackage{graphicx}
\usepackage{hyperref}
\usepackage{txfonts}
\newcommand{\punc}{\textit{Punctum}}
\begin{document}

   \title{ALMA discovery of \textit{Punctum} -- a highly polarized mm source in nuclear starburst galaxy NGC~4945}


   \author{E.~Shablovinskaia\inst{1}
          \and
          C.~Ricci\inst{1,2}
          \and 
          C-S.~Chang\inst{3}
          \and
          R.~Paladino\inst{4}
          \and
          Y.~Diaz\inst{1}
          \and 
          D.~Belfiori\inst{4}
          \and 
          S.~Aalto\inst{5}
          \and 
          M.~Koss\inst{6,7}
          \and 
          T.~Kawamuro\inst{8}
          \and
          E.~Lopez-Rodriguez\inst{9}
          \and 
          R.~Mushotzky\inst{10,11}
          \and 
          G.~C.~Privon\inst{12,13,14}}

   \institute{Instituto de Estudios Astrofísicos, Facultad de Ingeniería y Ciencias, Universidad Diego Portales, Av. Ejército Libertador 441, Santiago, Chile\\
              \email{elena.shablovinskaia@mail.udp.cl}
    \and Kavli Institute for Astronomy and Astrophysics, Peking University, Beijing 100871, China
    \and Joint ALMA Observatory, Avenida Alonso de Cordova 3107, Vitacura, Santiago 7630355, Chile
    \and INAF, Istituto di Radioastronomia di Bologna, via Piero Gobetti 101, 40129 Bologna, Italy
    \and Department of Space, Earth and Environment, Chalmers University of Technology, SE-412 96 Gothenburg, Sweden
    \and Eureka Scientific, 2452 Delmer Street Suite 100, Oakland, CA 94602-3017, USA
    \and Space Science Institute, 4750 Walnut Street, Suite 205, Boulder, Colorado 80301, USA
    \and Department of Earth and Space Science, Osaka University, 1-1 Machikaneyama, Toyonaka 560-0043, Osaka, Japan
    \and Kavli Institute for Particle Astrophysics \& Cosmology (KIPAC), Stanford University, Stanford, CA 94305, USA
    \and Department of Astronomy, University of Maryland, College Park, MD 20742, USA 
    \and Joint Space-Science Institute, University of Maryland, College Park, MD 20742, USA
    \and National Radio Astronomy Observatory, Charlottesville, VA 22903, USA
    \and Department of Astronomy, University of Florida, Gainesville, FL, 32611, USA
    \and Department of Astronomy, University of Virginia, Charlottesville, VA, 22904, USA}

   \date{Received September 15, 1996; accepted March 16, 1997}

 \abstract
{We report the discovery of a highly polarized millimeter (mm) continuum source in the central region of NGC~4945, identified through ALMA Band~3 observations. This starburst Seyfert 2 galaxy contains numerous compact mm sources, yet only one -- located approximately 3\farcs4 ($\sim$60~pc) from the galactic center and unresolved with $\sim$0\farcs1 resolution -- exhibits an unusually high polarization degree of 50\%~$\pm$~14\%, likely originating from non-thermal synchrotron radiation. The source is faint, yet clearly detected in two separate epochs of observation taken 14 days apart, with flux of 0.104~$\pm$~0.018 and 0.125~$\pm$~0.016~mJy, as well as in earlier ALMA observations, showing no variability at any timescale. The spectral index remains stable within large uncertainties, $-$1.8~$\pm$~2.5 and $-$1.3~$\pm$~2.5.
The source, which we further refer to as \textit{Punctum} due to its compactness, revealed no clear counterparts in existing X-ray or radio observations. Assuming association with the central region of NGC~4945, we estimate upper limits for its luminosity of $\sim$1~$\times$~10$^{37}$~erg~s$^{-1}$ in the 3--6~keV X-ray band (from archival \textit{Chandra} data) and $\sim$5~$\times$~10$^{35}$~erg~s$^{-1}$ at 23~GHz (from archival ATCA data).
A comparison of the radio, mm (including polarization), and X-ray properties with known astrophysical sources emitting synchrotron radiation, such as accreting neutron stars, supernova remnants, and non-thermal galactic filaments, revealed no clear match in any of these scenarios. The exact nature of this highly polarized source remains undetermined.}


   \maketitle
%

\section{Introduction}

Millimeter-wave (mm) observations in the $\sim$30--300~GHz range have become an important tool in extragalactic astronomy, offering unique access to probe cold gas, dust, and magnetic fields of galaxies. They enable us to trace star formation across cosmic time \citep[e.g.][]{decarli19,hodge20}, to map the dust and gas dynamics in both local \citep[e.g.][]{leroy21} and high-redshift galaxies \citep[e.g.][]{smit18,parlanti23}, and to reveal the large-scale structure of the magnetic field in galaxies even well beyond cosmic noon \citep{9io9,MFz56}. This progress has been made possible largely thanks to the Atacama Large Millimeter/submillimeter Array (ALMA), whose unprecedented combination of sensitivity and angular resolution has transformed our ability to resolve both compact and extended emission on sub-arcsecond scales.

Observations in mm band have also opened a window into the non-thermal emission produced by plasma in the immediate vicinity of accreting supermassive black holes (SMBHs) in active galactic nuclei (AGNs). The most striking demonstration of this capability is the imaging of the SMBH shadow and the toroidal magnetic field structure in the jetted AGN M87* by the \citet{EHT19,eht21}. While this is the most prominent example, it is not unique: even in non-jetted (so-called radio-quiet, RQ) AGNs, high-resolution mm observations have revealed compact cores whose fluxes correlate tightly with X-ray emission \citep{laor08,behar18,Kawamuro22,ricci23} and are consistent with synchrotron radiation from regions only a few tens of gravitational radii in size \citep{shab24,delpalacio25,rybak25}. The physical origin of this emission, potentially linked to the X-ray corona, compact jets, or shock-related processes, remains under active investigation.

To investigate the properties of synchrotron mm emission from compact cores in RQ AGNs, we initiated the first dedicated ALMA polarimetric campaign, with results detailed in \citet{shab25}. Our sample, drawn from the list of compact mm-bright RQ AGNs compiled by \citet{ricci23}, included NGC~4945, a well-known nearby galaxy at a redshift of $z = 0.0023$ and a distance of $D = 3.72$~Mpc\footnote{The redshift-independent distance was measured using the tip of the red giant branch (see \citealp{koss22}) and is consistent with the measurements from \cite{karachentsev07}.}. Viewed nearly edge-on, NGC~4945 hosts a heavily obscured, Compton-thick AGN with a column density of $\log N_{\rm H} = 24.80$~cm$^{-2}$ \citep{ricci15,ricci17}, and is optically classified as a type~2 AGN \citep{Lipo88}. Alongside its active nucleus detected at 100--220~GHz \citep{kawamuro23,ricci23}, the galaxy exhibits intense nuclear star formation, as revealed by free–free continuum and hydrogen recombination line emission \citep{emig20}. This makes NGC~4945 a particularly attractive target not only for probing the polarized mm emission from its AGN core (which was found to be undetected at the 0.5\% level; \citealt{shab25}), but also for investigating the polarization properties of other mm sources throughout the galaxy.

In this paper, we present ALMA Band~3 polarization observations of the $\sim$200~pc inner region of the active starburst galaxy NGC~4945. Our findings indicate that nearly all mm-emitting sources in the galaxy show no detectable polarization (much less than 1.5\%), with the exception of a peculiar unidentified source located above the galactic plane, which exhibits extremely high polarization. In the following sections, we summarize the properties of this source and compare them with the potential origins of polarized mm emission.  Throughout the paper we used the standard cosmological parameters ($H_0 = 70$~km\,s$^{-1}$\,Mpc$^{-1}$, $\Omega_{\rm m} = 0.3$, $\Omega_\Lambda = 0.7$).

\section{Observations and data reduction}

The ALMA Band~3 observations of NGC~4945 were conducted in two sessions: the first on 2023$-$10$-$11 and the second on 2023$-$10$-$25 (project code 2023.1.01517.S; PI C. Ricci). The spectral setup was in Time-Division Mode (TDM) centered around 100~GHz, targeting the continuum emission of the galaxy. The observations were performed across four spectral windows with a 1.985~GHz bandwidth and central frequencies of 90.52, 92.48, 102.52, and 104.48~GHz. During the first session (hereafter referred to as S1), an insufficient parallactic angle coverage (51.12$^\circ$ instead of the desired 60.00$^\circ$) for the polarization calibrator J1256--0547 prompted the need for a second session (hereafter S2).

Observations were scheduled during the ALMA long baseline configuration (C-8), with the longest baseline extending to 8.5~km. The restored beam size for S1 was 0\farcs123 $\times$ 0\farcs105 (2.2 $\times$ 1.9~pc), and for S2 it was 0\farcs140 $\times$ 0\farcs102 (2.5 $\times$ 1.8~pc). The on-source integration times were 38\,minutes for S1 and 75\,minutes for S2. The rms background in the restored ALMA image in total intensity is $\sigma$ = 0.017~mJy/beam for S1 and 0.014~mJy/beam for S2. Data processing was carried out using \texttt{CASA} version 6.5.4.9 and ALMA Pipeline version 2023.1.0.124 \citep{alma_pipeline}. The clean images were generated using CASA task \texttt{tclean} with weighting = briggs (robust=0.5).

According to the ALMA Proposer’s Guide, the systematic flux error for Band~3 observations is 5\%. Flux measurements from the polarization calibrators J1256$-$0547 (in S1) and J1427$-$4206 (in S2), as well as the check source J1254$-$4743, yielded consistent results, justifying the application of a 5\% error margin in our analysis. Furthermore, the estimated degree and angle of polarization for the calibrators were consistent with the measurements from the AMAPOLA calibrator monitoring\footnote{\url{https://www.alma.cl/~skameno/AMAPOLA/}} at both the S1 and S2 epochs, indicating that the polarization measurements at S1 also remain reliable.

For each epoch, we created maps of the total and polarized intensity. The circular polarization component was assumed to be zero, as it is typically weak in the mm regime, so the polarized intensity was calculated as $P = \sqrt{Q^2 + U^2}$, where $Q$ and $U$ are the Stokes parameters in units of flux. The polarization degree was then calculated as PD = $P/I$, where $I$ is the total intensity, and debiased following the method outlined in \citet{montier1,montier2}. The total and polarized intensity maps, as well as the PD maps for the selected sources, are shown in Fig.~\ref{fig:big} and Fig.~\ref{fig:app}.

\section{Results}

\begin{figure*}
    \centering
    \includegraphics[width=0.95\linewidth]{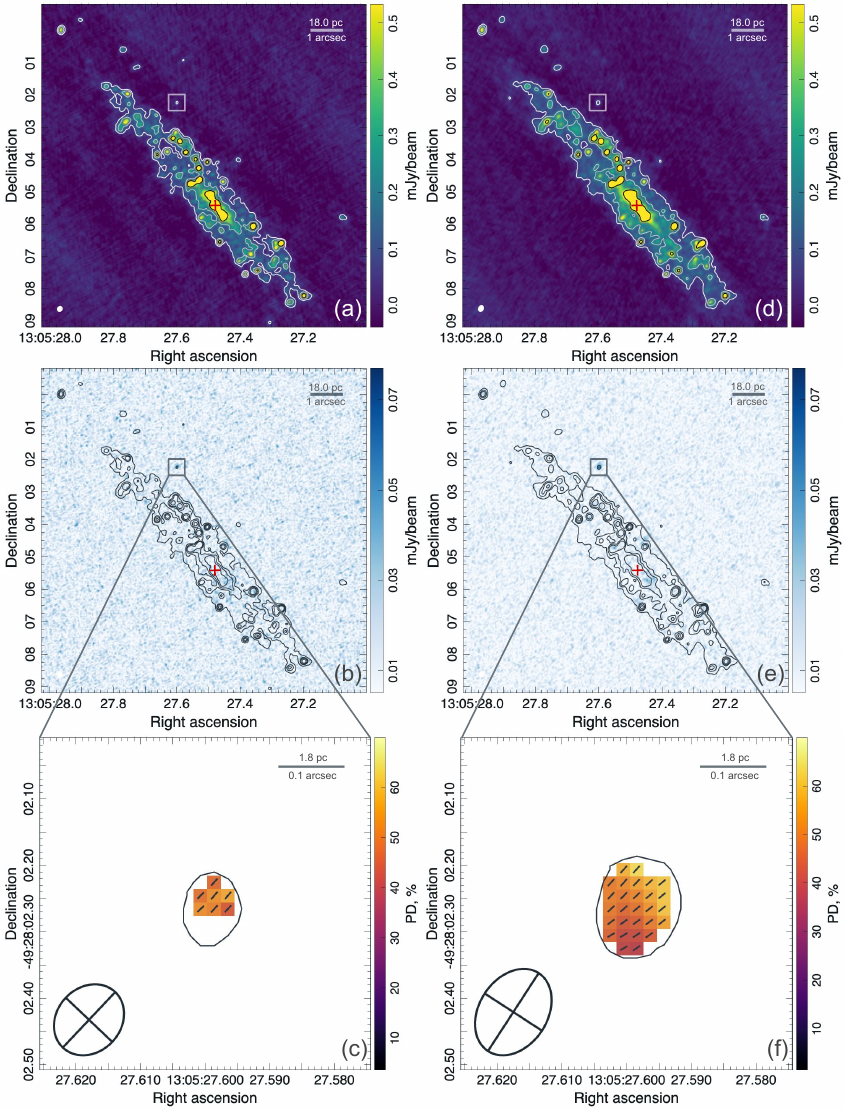}
    \caption{ALMA observations of NGC~4945 on 2023$-$10$-$11 (S1, left column) and on 2023$-$10$-$25 (S2, right column). \textit{Upper panel}: total intensity with the contours corresponding to the $0.07, 0.19, 0.31, 0.54$~mJy/beam levels in the box $\sim$10$\times$10$''$. The location of the AGN is marked with a red cross. \textit{Middle panel}: polarized intensity with the overplotted total intensity contours. \textit{Bottom panel}: polarization degree in per cent in the box of $\sim$0.5$\times$0.5$''$. The orientation of the polarization vector is given with black ticks.}
    \label{fig:big}
\end{figure*}

Despite the abundance of bright mm sources in the central part of NGC\,4945, as seen in the total intensity maps in Fig.\ref{fig:big}a,d, nearly all sources disappear in polarized light. The polarized intensity maps (Fig.~\ref{fig:big}b,e) reveal only a few regions with significant polarized mm emission. Note here that the central mm emission in NGC~4945 spans about 14$''$, covering less than one-third of the $\sim$60$''$ field of view, which ensures accurate polarization calibration \citep{cortes_2023_7822943}. To accurately estimate the polarization degree, we created the PD maps masking areas with a signal-to-noise ratio (S/N) $<$ 5 to avoid bias and unreliable polarization measurements in low-S/N regions. Initially, for both observation epochs, S1 and S2, we identified several compact regions with polarization exceeding $\sim$1.5\%. However, upon comparing these maps, we found that most of these initially detected polarized regions did not appear in both epochs, preventing us from confirming them as real sources. Ultimately, we confirmed only two polarized regions that were consistently observed during both S1 and S2.

The first polarized region is an extended structure with a flux of $\sim$0.5~mJy, located 0\farcs5 (i.e. $\sim$9~pc) from the central AGN (RA 13:05:27.440, Dec $-$49:28:05.670). Maps of polarized intensity and PD are shown in Fig.~\ref{fig:app}. This source resides at the edge of the bright molecular disk, which, based on our ALMA Band~3 observations, reprocessed into spectral cubes with a channel width of 31.2~MHz, emits in both continuum and several emission lines, including HCN(1--0), HC$_3$N(10--9), and N$_2$H$^+$(1--0). Due to its proximity to a much brighter unpolarized mm emitting extended region, it is difficult to isolate the source of polarized emission. Contamination from the unpolarized component could significantly reduce the measured polarization. The estimated polarization degrees are 14\%~$\pm$~4\% and 8\%~$\pm$~2\% for S1 and S2, respectively, with errors given at the 3$\sigma$ level.  Given the difficulty in isolating this region from brighter sources, we do not focus on it further; its PD map is included in Appendix~\ref{appendA}.

The most intriguing structure was found $\sim$3\farcs4 (i.e. $\sim$61~pc) away from the galactic center (RA 13:05:27.600, Dec $-$49:28:02.242). This compact source is well isolated from other mm continuum sources in the galactic disk, despite being faint. We measured its flux in each epoch, finding values of 0.104~$\pm$~0.018~mJy for S1 and 0.125~$\pm$~0.016~mJy for S2 averaged over all spectral windows with S/N~$\approx$~6.1 and 8.9, respectively. Over the 14-day interval between the two epochs, the flux of this polarized structure varied by $\sim$18\%, corresponding to a difference of only $\sim$1.2$\sigma$ -- not statistically significant, though potentially suggestive of variability. Additionally, we measured the flux in each of the four spectral windows (see Table~\ref{tab:flux}), finding no statistically significant variations between epochs. The spectral indices $\alpha$, obtained by fitting the spectra with a power law ($\propto \nu^{\alpha}$) and accounting for flux uncertainties, were $\alpha = -1.8 \pm 2.5$ for S1 and $\alpha = -1.3 \pm 2.4$ for S2, showing no statistically significant difference between the two epochs. All values are summarized in Table~\ref{tab:flux}; note that the flux uncertainties were computed as the quadrature sum of the 5\% Band~3 calibration uncertainty and the background rms measured in each corresponding image. 

Furthermore, we analyzed the source’s flux in each spectral window, dividing it into frequency bins with a 31.2~MHz step. Due to the faintness of the source, this spectral analysis mostly revealed noise, with no evidence of emission lines typically found in the molecular gas of NGC~4945, not even at the 1$\sigma$ level.

\begin{table}[!ht]
\centering
\caption{ALMA Band~3 fluxes and the spectral index of the highly polarized mm source in NGC 4945 for epochs S1 (2023--10--11) and S2 (2023--10--25).}
\begin{tabular}{ccc}
\hline
Frequency & Flux S1 & Flux S2 \\
(GHz) & (mJy) & (mJy) \\
      \hline
total  & 0.104 $\pm$ 0.018 & 0.125    $\pm$ 0.016    \\
\hline
90.5      & 0.110  $\pm$ 0.035      & 0.132   $\pm$ 0.028       \\
92.5      & 0.098  $\pm$ 0.036      & 0.142   $\pm$ 0.028       \\
102.5     & 0.097  $\pm$ 0.040      & 0.117   $\pm$ 0.029       \\
104.5     & 0.082  $\pm$ 0.037      & 0.101  $\pm$ 0.030        \\
\hline
$\alpha$   & $-$1.8 $\pm$ 2.5 & $-$1.3 $\pm$ 2.4 \\
\hline
\end{tabular}
\label{tab:flux}
\end{table}

Although faint in total flux, the source became the brightest in polarized intensity across NGC~4945. In the PD map, the source exhibits an exceptionally high level of polarization, consistent across both epochs: 51\%~$\pm$~14\% for S1 and 50\%~$\pm$~14\% for S2, with the polarization angle of 133.2$^\circ$~$\pm$~7.9$^\circ$ for S1 and 130.6$^\circ$~$\pm$~7.8$^\circ$ for S2. Unfortunately, the source is too faint to reliably estimate the PD per frequency or per spectral window due to increased noise, preventing us from tracing polarization changes across the spectral windows.

\section{Search for the counterparts}

\begin{figure}
    \centering
    \includegraphics[width=0.95\linewidth]{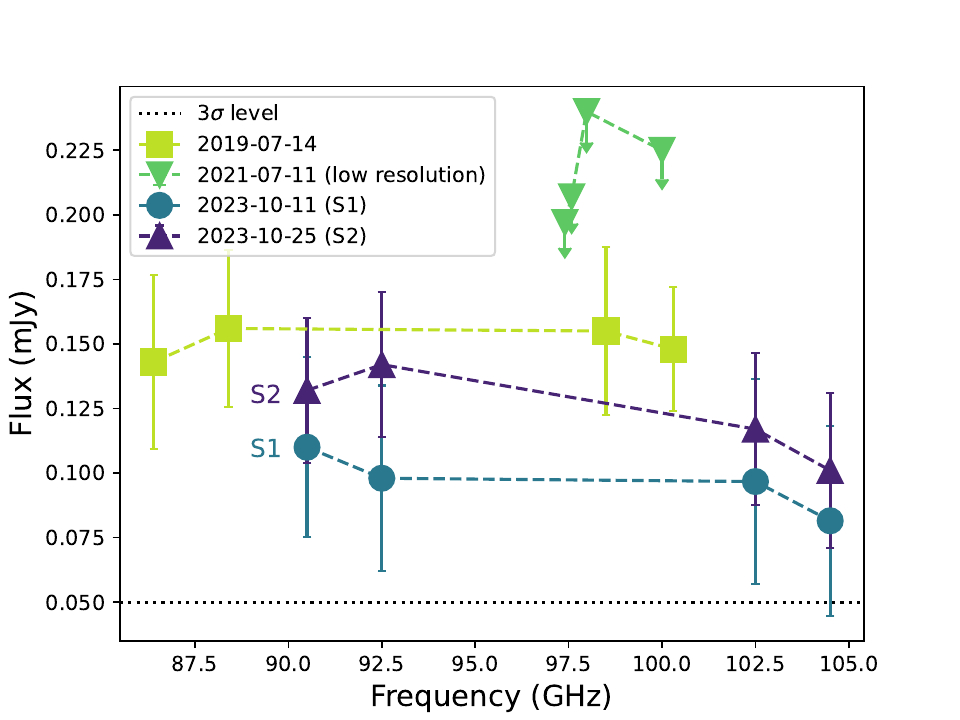}
    \caption{ALMA observations of the \punc\ in different epochs. Due to the low resolution in 2021--07--11 epoch, only upper limits are given. The 3$\sigma \approx 0.05$~mJy level for S1 and S2 epochs is given in a black dotted line.}
    \label{fig:sed}
\end{figure}

It is quite rare to observe such a large PD -- about 50\% -- in astrophysical sources.  This significant PD, combined with the steep spectral index, strongly suggests that the emission can only be explained by an optically thin synchrotron source with a highly uniform magnetic field structure. To better understand the nature of this source, which we refer to as \punc\footnote{From Latin \textit{pūnctum}, meaning "point" or "dot".} due to its compactness and unknown origin, observations at other wavelengths should be explored.

\noindent\textbf{Radio.} According to ATCA data from \citet{lenc2009}, no specific or isolated source corresponding to \punc\ was detected at frequencies of 17, 19, 21, and 23~GHz, with 1$\sigma$ detection limits of 0.25--0.34~mJy/beam at a resolution of $\sim$0\farcs5. In the same study, high-resolution ($\sim$tens of mas) LBA data at 2.3~GHz also show no detection of the source with a detection limit of $\sim$0.1~mJy/beam. Based on the 23~GHz non-detection, we can place a rough upper limit of $\sim$5~$\times$~10$^{35}$~erg~s$^{-1}$ on the radio luminosity of \punc\ at this frequency. No specific detections were made with VLBA in the water maser lines either \citep{Greenhill1997}.
In our ALMA Band~3 dataset, the flux density and spectral index uncertainties are so large that a reliable radio flux extrapolation is impossible. As a result, it remains unclear whether the radio upper limits are consistent with the mm data or whether the mm emission is instead affected by absorption effects or source variability.

\noindent\textbf{Other mm bands.} When the resolution is worse than 0\farcs3, \punc\ cannot be separated from the mm emission of other nearby structures. According to the ALMA archive, observations with a resolution better than 0\farcs3 were conducted in Bands 3, 7, and 8. In Band 7 ($\sim$350~GHz), no detection was made above the 3$\sigma$ = 0.3~mJy/beam level. In Band 8, there is a faint detection of a structure with a flux of approximately 4.5~mJy, though it is comparable to the noise level, with 1$\sigma$ = 3.5\,mJy. No molecular emission was detected in this region \citep{Henkel2018}, nor were any water megamasers at 183~GHz \citep{Humphreys2016}. 

While NGC~4945 has been observed multiple times in Band~3 with high resolution, only two additional epochs, apart from our observations, had sufficient on-source time to detect the source at least at the 3$\sigma$ level. The archival Band 3 data, together with our new observations, are presented in Fig.~\ref{fig:sed}. For the epoch 2021--07--11 (2019.1.00416.S, PI: He), the relatively low resolution ($\sim$0\farcs3) allowed only upper limits to be derived for the flux. However, for the 2019--07--14 epoch (2018.1.01236.S, PI: Leroy), the resolution ($\sim$0\farcs1) matched that of S1 and S2. In this case, the flux averaged over the spectral windows was measured to be 0.150~$\pm$~0.015~mJy, assuming a 5\% error and background rms $\sigma = 0.014$~mJy/beam, with S/N~$\approx$~11. This indicates that the observed flux variations of \punc\ over the four-year period remain within 3$\sigma$ uncertainties and are therefore not statistically significant. It is also noteworthy that, for this epoch, the spectral slope is flatter, $0.02 \pm 1.51$, but comparable to that observed during polarimetric observations within uncertainties. 

\noindent\textbf{IR and optical.} For NGC~4945, data from the \textit{James Webb Space Telescope} are available for other regions of the galaxy, though not for the zone of interest. \textit{Spitzer} observations of the central region of NGC~4945 have saturated flux, making them unsuitable for further analysis. Concerning the optical observations, NGC~4945 is oriented almost edge-on and is a type\,2 AGN, in which the central parts are covered with dust. Therefore, no specific features can be found in archival \textit{HST} data from 1999--2000, as well as in more recent MUSE observations from 2015 (094.B-0321, PI: Marconi), which show no specific features in this area in the $\sim$4760--9340\AA{} range.

\noindent\textbf{X-ray.} For NGC~4945, several epochs of \textit{Chandra} observations are available. Given the faintness of the object, we selected the 2013--04--25  epoch with the longest exposure of 130.5~ksec (ObsID 14984). The central region of NGC~4945 exhibits extended emission ($\sim$10$''$), which is associated with the central AGN and the surrounding star-forming region. However, we did not identify any specific counterpart to \punc.

To estimate the upper limit of the source's X-ray luminosity, we extracted the X-ray spectra and applied a \texttt{phabs*powerlaw} model, where \texttt{phabs} represents the Galactic absorption predicted using the NH tool within FTOOLS \citep{Dickey90,LAB}. We then used the cflux model of XSPEC to calculate the flux in the 3.0--6.0 keV band, resulting in an upper limit of 7.1~$\times$~10$^{-15}$~erg~s$^{-1}$~cm$^{-2}$ with a photon index of $-$0.60$^{+1.07}_{-0.09}$, corresponding to a luminosity of 1.2 $\times$ 10$^{37}$~erg~s$^{-1}$.
It is important to note that this represents an upper limit, as contributions from surrounding star-forming material were not accounted for. No distinct X-ray source was identified as a counterpart to the polarized mm emission.

\section{Discussion}

In summary, we observe an unresolved source with a mm flux of $\sim$0.1~mJy at 100~GHz and an exceptionally high polarization of $\sim$50\%. The absence of an optical counterpart suggests that the source is not Galactic and is most likely embedded within the central region of the galaxy NGC~4945, obscured by dust. Furthermore, the source is unlikely to be a background object, such as a blazar, which are known to exhibit much lower degrees of polarization in the mm band, typically $\sim$15\% at maximum, according to the AMAPOLA database\footnote{\url{https://www.alma.cl/~skameno/AMAPOLA/}}. 
Moreover, while the number density of blazars at such low flux levels is not commonly reported, we extrapolated the number counts from \citet{tucci11} and estimated that the probability of finding a blazar with a flux of $\sim$0.1~mJy within 3\farcs4 of the nucleus of NGC~4945 is less than 0.4\%.
Based on this, in the following discussion, we will assume that the source is located in NGC~4945.

\punc\ was detected across multiple epochs of high-resolution ALMA observations (see Fig.~\ref{fig:sed}), showing no significant flux variations beyond the 3$\sigma$ level between 2019 and 2023. This effectively rules out a transient event such as a gamma-ray burst (GRB), as the mm flux from GRBs typically diminishes rapidly within the first two hundred days \citep{grb_lc}, and their mm polarization is expected to be very low (<1\%, \citealt{grb}). The observed mm polarization is also significantly higher than that typically associated with sources in which polarization arises from scattering on dust, as seen in protoplanetary disks, which exhibit only a few percent of polarization \citep{alma_pp}. Similarly, polarization from magnetized thermal dust is much lower, ranging from $\lesssim$1\% \citep[e.g.][]{9io9} to $\sim$3--4\% \citep[e.g.][]{bonavera17,Clements25}. Moreover, the expected spectral index of dust emission in the mm regime is $\sim$3.5 \citep{mullaney11}, which definitively rules out a dust-related origin.

The polarization of $\sim$50\% suggests the presence of an ordered magnetic field. As only continuum emission has been detected from this source, with no emission lines present, a maser origin is ruled out. The most plausible scenario is the synchrotron emission of a source with an ordered magnetic field. The source has significant luminosity in the mm band, $L_{\rm mm} \approx 2\times 10^{35}$~erg~s$^{-1}$\footnote{This estimate assumes isotropic mm emission. While the nature of the source remains uncertain, relativistic beaming cannot be excluded and may imply a lower intrinsic luminosity.}. Moreover, the source is expected to be compact. As in ALMA observations, its structure was unresolved, and the size should be $\lesssim$2~pc. However, since no significant variability was detected between epochs, we cannot place any upper limit on the source size based on variability timescales. Additionally, the large uncertainties in the spectral index across all epochs in both 2019 and 2023 prevent us from reliably identifying a spectral turnover, making it difficult to constrain the source size via synchrotron self-absorption modeling. Assuming that the spectrum observed on 2019--07--14 is indeed flat due to synchrotron self-absorption, the source size can be estimated using Eq.~19 from \citet{laor08}, yielding 2 $\times$ 10$^{-6}$~pc ($\sim$10$^{13}$~cm) for magnetic field of 1~Gauss and 4 $\times$ 10$^{-7}$~pc ($\sim$10$^{12}$~cm) for 1~mGauss. These estimates, however, remain highly uncertain, and in reality, the size of the emitting region is still largely unconstrained.

Among the known types of sources that can produce such high polarization degrees, accreting neutron stars are suitable candidates. Indeed, the \punc's spectral index of $-1.5$~$\pm$~1.7, calculated as the weighted mean from two epochs, aligns with typical pulsar spectral indices, with a mean value of $\alpha = -1.8$ \citep{maron00}. However, as noted by \citet{torne17}, pulsars are challenging to detect at radio frequencies above a few gigahertz, with only a handful of detections above 30~GHz \citep[see][for review]{torne17}. The pulsar with the highest flux density detected so far by ALMA, the Vela pulsar, was observed at 97.5 to 343.5~GHz and exhibited a flux of just $\sim$0.252~mJy at 97.5~GHz \citep{vela_alma}. Assuming its distance $\sim$300~pc \citep{caraveo01}, the Vela pulsar luminosity is $\sim$3 $\times$ 10$^{27}$~erg~s$^{-1}$, which is $10^8$ times fainter than our source. The other mm emitting pulsar, PSR~J1023+0038, was found to be 10 times brighter, i.e., showing the luminosity $\sim$3~$\times$~10$^{28}$~erg~s$^{-1}$ \citep{baglio_alma_pulsar}, which is still too faint to explain the emission observed in our case. 

Magnetars, a subclass of neutron stars, are comparatively bright in the mm band and belong to the top 3\% of the most luminous pulsars, characterized by stronger magnetic fields \citep{torne18}. Notably, high polarization degrees have been observed in magnetars. For example, \citet{liu21} reported nearly 100\% linear polarization at 86~GHz in SGR~J1745--2900, albeit just during the pulse phases. Additionally, \citet{pulsar_7mm} observed 65\% of the averaged polarization of the same source at 40--48~GHz with the VLA, though no similar measurements have been made at higher frequencies.

Despite the high polarization supporting a magnetar interpretation, the mm luminosities of known magnetars are too faint to match the observed source. Several magnetars have been observed at high frequencies: XTE J1810--197 ($D \approx 3.5$~kpc; \citealt{Minter08}) showed 1.2~mJy at 88~GHz \citep{camilo07}, corresponding to $\sim$2 $\times$~10$^{30}$~erg~s$^{-1}$, while SGR J1745–2900, orbiting Sgr~A* ($D \approx 8$~kpc; \citealt{sgr_dist}), exhibited fluxes between 0.6 and 6.4~mJy at 101~GHz \citep{torne15, torne17}, equivalent to $\sim$5 $\times$~10$^{31}$~erg~s$^{-1}$, resulting in a substantial luminosity gap of a factor of 10,000 to 100,000 between these known sources and \punc.

The only neutron-star-associated sources with comparable mm luminosities are supernova remnants (SNRs). Given the calculated star-formation rate of $\sim$2--8~$M_\odot$~yr$^{-1}$ for NGC~4945 \citep{roy10}, we can estimate the supernova rate using the relation from \citet{Cappellaro99}, yielding $\sim$2--8 SN per 100 years. This suggests that young supernova remnants are highly probable in this galaxy. Indeed, two supernovae have been detected in NGC~4945 in recent years -- SN 2005af \citep{2005af} and 2011ja \citep{2011ja}. However, both are located far from the central region, at $\sim$5 and 10~kpc, respectively, and no recent SN has been detected in the vicinity of \punc. This absence could, however, be attributed to significant dust obscuration, which may prevent the optical identification of a young supernova in the central regions of the galaxy.

SNR can provide a high mm luminosity. For example, SN~1987A ($D \approx 51.2$~kpc; \citealt{sn1987a_dist}) shows flux in the range of $\sim$10--20~mJy at 94~GHz  \citep{sn1987a_mm}, corresponding to $\sim$6~$\times$~10$^{33}$~erg~s$^{-1}$. The brightest SNR, the Crab Nebula (M1), exhibits a flux of $\sim$300~Jy at $\sim$100~GHz at a distance of 2~kpc \citep{crab_nebula}, equivalent to $\sim 2 \times 10^{35}$~erg~s$^{-1}$. Moreover, the typical X-ray luminosities of SNRs are consistent with the upper limit of the X-ray luminosity observed for \punc. X-ray observations of SNRs in the LMC \citep{SNR_LMC} reveal luminosities ranging from $10^{34}$ to $10^{37}$~erg~s$^{-1}$. The radio-to-mm spectral index of M1 is nearly flat (around $-0.3$), which, within the uncertainties, could be consistent with that of \punc. However, the mm polarization of M1 is significantly lower ($\sim$7--8\%; \citealt{crab_pol}), and thus does not fully align with the high polarization observed in \punc.

Microquasars, stellar-mass black holes or neutron stars, could be potential candidates, as their jets can produce bright mm emission with luminosities comparable to those of SNRs, reaching up to $\sim$$10^{33}$~erg~s$^{-1}$ \citep[e.g.,][]{ss433}. In certain states, their radio spectra become self-absorbed at low frequencies, with luminosities of $10^{29}$--$10^{32}$~erg~s$^{-1}$ at frequencies of a few tens of GHz \citep[e.g.,][]{fender01}, which is broadly consistent with the upper limit derived for \punc.  However, \citet{ss433} reported very low mm polarization (<1\%) in the brightest microquasar SS~433, which contradicts our observations.

Synchrotron emission can also arise from non-thermal filamentary structures located a few tens to hundreds of parsecs away from the galactic center. In the Milky Way galaxy, non-thermal filaments and radio loops were first discovered in the radio band (1.5~GHz, \citealt{Yusef84}) as long, narrow, and highly collimated structures. Subsequent observations at various radio frequencies \citep[e.g.,][]{Heywood22,vidal15} revealed steep radio spectra with spectral indices around $-1$, consistent with the synchrotron origin of the filaments' radiation. \citet{vidal15} demonstrated that, while the median radio polarization of the filaments is $\sim$10\%, certain regions exhibit linear polarization as high as $\sim$40\%. This aligns with earlier findings by \citet{gray95}, who reported up to 50\% polarization in a segment of one filament. However, despite the observed polarization being consistent with the mm polarization of our polarized source, the brightest filaments have a maximum luminosity on the order of $10^{33}$~erg~s$^{-1}$ at 5~GHz \citep{gray95}. Given their steep spectra, the luminosity at 100~GHz would decrease to only a few times $10^{32}$~erg~s$^{-1}$, which remains insufficient to account for our mm source.

To sum up, none of the currently known sources exhibit the combination of properties -- most notably the mm luminosity and polarization -- seen in \textit{Punctum}, highlighting its distinct nature.

\section{Summary}

The unprecedented sensitivity and angular resolution of ALMA, including its polarimetric capabilities, enabled deep polarimetric observations of the central starburst region in the type~2 active galaxy NGC~4945. In these observations, we discovered a faint ($\sim$0.1~mJy), compact millimeter source located 3.4~pc from the AGN. In total intensity, the object, which we called \punc, resembles other compact mm sources in NGC~4945, associated with embedded star clusters. However, unlike the others, it exhibits significant linear polarization at a level of 50\%~$\pm$~14\%. Assuming it resides within the central region of NGC~4945, we estimate its mm luminosity to be $\sim$2~$\times$~10$^{35}$~erg~s$^{-1}$. Based on archival \textit{Chandra} and ATCA data, we place upper limits on its X-ray and radio luminosities of $1$~$\times$~10$^{37}$ and $5$~$\times$~10$^{35}$~erg~s$^{-1}$, respectively. Unfortunately, \punc\ is too faint to reliably measure a spectral index or assess variability. As it remains unresolved, we place an upper limit of $\sim$2~pc on its size.

The most plausible mechanism for producing such a high degree of linear polarization is synchrotron emission. We compared the properties of \punc\ with known classes of synchrotron-emitting sources. Among these, magnetars are the most likely candidates due to their high polarization, though their mm luminosities are generally lower. SNRs, which have comparable mm and X-ray luminosities, typically exhibit low polarization. Galactic non-thermal filaments can produce high polarization, but are usually faint in the mm regime. Thus, \punc\ cannot currently be associated with any known class of astrophysical source.

To determine its nature, deeper polarimetric observations in both mm and radio bands are required to constrain the spectral index and investigate potential frequency dependence of the polarization. High-resolution, multi-wavelength follow-up observations would also be invaluable for further characterization.

\begin{acknowledgements}
 This paper makes use of the following ALMA data: ADS/JAO.ALMA\#2023.1.01517.S. ALMA is a partnership of ESO (representing its member states), NSF (USA) and NINS (Japan), together with NRC (Canada), NSTC and ASIAA (Taiwan), and KASI (Republic of Korea), in cooperation with the Republic of Chile. The Joint ALMA Observatory is operated by ESO, AUI/NRAO and NAOJ.
 We sincerely thank Ari Laor for the valuable comments and feedback.
 ES acknowledges ANID BASAL project FB210003 and Gemini ANID ASTRO21-0003. CR acknowledges support from Fondecyt Regular grant 1230345, ANID BASAL project FB210003 and the China-Chile joint research fund.
 YD acknowledges the financial support from a Fondecyt postdoctoral fellowship (3230310). TK is supported by JSPS KAKENHI grant numbers 23K13153/24K00673.
\end{acknowledgements}

   \bibliographystyle{aa} 
   \bibliography{bib} 
%

\begin{appendix}

\onecolumn
\section{Region of moderate mm polarization} \label{appendA}

\begin{figure*}[h!]
    \centering
    \includegraphics[width=0.95\linewidth]{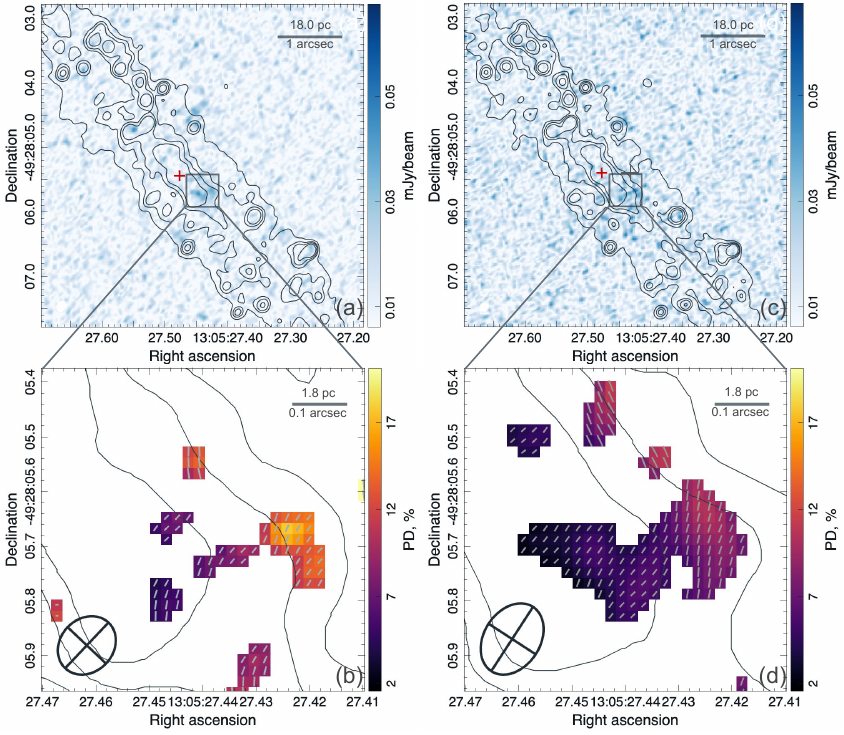}
    \caption{ALMA observations of NGC~4945 on 2023--10--11 (S1, left column) and on 2023--10--25 (S2, right column). \textit{Upper panel}: polarized intensity with the overplotted total intensity contours corresponding to the $0.07, 0.19, 0.31, 0.54$~mJy/beam levels in the box $\sim$5$\times$5$''$. \textit{Bottom panel}: polarization degree in per cent in the box of $\sim$0.6$\times$0.6$''$. The orientation of the polarization vector is given with gray ticks.}
    \label{fig:app}
\end{figure*}

\end{appendix}

\end{document}